\documentclass[aps,prd,superscriptaddress,twocolumn]{revtex4}

\usepackage{amsfonts}
\usepackage{amsmath}
\usepackage{graphicx}
\usepackage{color}
\usepackage{verbatim}
\usepackage[dvips]{epsfig}
\usepackage{float}
\usepackage{epsfig,subfig}
\usepackage{epstopdf}
\usepackage{hyperref}
\usepackage{cleveref}

\DeclareMathAlphabet\mathbfcal{OMS}{cmsy}{b}{n}

\begin{document}

\title{Casimir effect in polymer scalar field theory}

\author{C. A. Escobar}
\affiliation{Instituto de F\'{i}sica, Universidad Nacional Aut\'{o}noma de M\'{e}xico, Apartado Postal 20-364, Ciudad de M\'{e}xico 01000, M\'{e}xico}

\author{E. Chan-L\'{o}pez}
\affiliation{Universidad Ju\'{a}rez Aut\'{o}noma de Tabasco, DACB, 86690 Cunduac\'{a}n, Tabasco, M\'{e}xico}

\author{A. Mart\'{i}n-Ruiz}
\email{alberto.martin@nucleares.unam.mx}
\affiliation{Instituto de Ciencias Nucleares, Universidad Nacional Aut\'{o}noma de M\'{e}xico, 04510 Ciudad de M\'{e}xico, M\'{e}xico}
\affiliation{Centro de Ciencias de la Complejidad, Universidad Nacional Aut\'{o}noma de M\'{e}xico, 04510 Ciudad de M\'{e}xico, M\'{e}xico}

\begin{abstract}
In this paper, we study the Casimir effect in the classical geometry of two parallel conducting plates, separated by a distance $L$, due to the presence of a minimal length $\lambda$ arising from a background independent (polymer) quantization scheme. To this end, we polymer-quantize the classical Klein-Gordon Hamiltonian for a massive scalar field confined between the plates and obtain the energy spectrum. The minimal length scale of the theory introduces a natural cutoff for the momenta in the plane parallel to the plates and a maximum number of discrete modes between the plates. The zero-point energy is calculated by summing over the modes, and by assuming $\lambda \ll L$, we expressed it as an expansion in powers of  $1/N$, being $N=L/ \lambda$ the number of points between the plates. Closed analytical expressions are obtained for the Casimir energy in the cases of small and large scalar mass limits.
\end{abstract}

\maketitle

\section{Introduction}

One of the most important challenges of modern theoretical physics is the search for a quantum theory of gravity (QTG). Theoretically, the major problem is that  the introduction of gravity into quantum field theories appears to spoil their renormalizability, and from the experimental side, the greatest difficulty is the lack of experimentally accessible phenomena that could shed light on a possible route to QTG. Because of these limitations, at present what physicists do is to look for quantum gravity effects through high-sensitivity measurements.

Polymer quantization (PQ) is a background independent quantization scheme inspired by Loop Quantum Gravity (LQG) \cite{Ashtekar} which has been used to explore mathematical and physical implications of theories such as quantum gravity. PQ may be viewed as a separate development in its own right, and is applicable to any classical theory whether or not it contains gravity. In short, its central feature is that the momentum operator $p$ is not realized directly as in Schr\"{o}dinger quantum mechanics because of a built-in notion of discreteness, but arises indirectly through the translation operator $U _{\lambda} = e ^{i \frac{p \lambda}{\hbar}}$, being $\lambda$ a fundamental length scale of the theory. Noticeably, while various approaches to QTG (such as LQG, String Theory and Non-Commutative Geometries) suggest the existence of a minimum measurable length \cite{Hossenfelder, Ali, Kempf}, in PQ a length scale is required for its construction.

PQ has been used to study quantum gravitational effects upon simple quantum systems, such as the harmonic oscillator \cite{Ashtekar2}, the particle in a box \cite{Corichi}, the diffraction in time \cite{AMR1}, the tunneling phenomena \cite{AMR2}, the Coulomb potential \cite{Husain, Kunstatter}, the quantum bouncer \cite{AMR3} and within statistical thermodynamics \cite{Chacon}, just to name a few. It has also been applied to the scalar field theory \cite{Ashtekar3, Hossain, Vergara, Kajuri} and the electromagnetic field \cite{Yuri}. In this work we develop the area of polymer quantum field theory further by computing the Casimir pressure between two parallel conducting plates in a polymer scenario.

In its simple manifestation, the Casimir effect is a quantum force of attraction between two parallel uncharged conducting plates \cite{Casimir}. More generally, it refers to the stress on bounding surfaces when a quantum field is confined to a finite volume of space \cite{Deutsch}. The boundaries can be material media, interfaces between two phases of the vacuum, or topologies of space-time. In any case, the modes of the quantum fields are restricted, giving rise to a macroscopically measurable force. For a review see, for example, Refs. \cite{Milton, Bordag}. The experimental accessibility to micrometer-size physics has motivated the theoretical study of the CE in different scenarios, including Lorentz-breaking extensions of the QED \cite{Kharlanov, AMR4, AMR5}, Lorentz-violating scalar \cite{Petrov1, Petrov2}  and fermionic \cite{Petrov3} field theory, topological phases of matter \cite{Cortijo1, Cortijo2, AMR6}, string theory \cite{Fabinger, Gies, Hadaz} and theories with minimal length based on Generalized Uncertainty Principles (GUPs) \cite{Ulrich, Nouicer, Frasino}.

One of the most commonly used procedures for obtaining the vacuum energy is direct evaluation of infinite sums over eigenvalues of zero-point field modes. These sums, which happen to be formally divergent, may be regularized by a variety of techniques, e.g., momentum cutoff or dimensional regularization \cite{Milton, Bordag}. An important feature about the boundary conditions is that in quantum models with minimal length there is a finite number of modes. So there is a natural cutoff and the Casimir energy does not need to be regularized, as opposed to the standard quantum field theory calculations. In other words, the minimal length acts as a regulator in the ultraviolet. Unlike the GUP scenarios, where String Theory predicts that it is impossible to improve the spatial resolution below the characteristic length of the string (and hence yields to modified position-momentum uncertainty relation to account for it) \cite{Townsend, Veneziano, Maggiore1, Maggiore2}, in PQ the minimal length is built-in by construction. Accordingly, the minimal length of the polymer theory must manifest in the Casimir effect. In this work, we aim to provide additional theoretical predictions about the quantum vacuum in polymer scalar field theory; in particular, we focus on the evaluation of the Casimir stress upon parallel uncharged conducting plates.

The outline of this paper is as follows. We begin in Sec. \ref{PolScalarField} by reviewing the polymer quantization scheme. In Sec. \ref{ConfinedPolScalField} we discuss a confined polymer-quantized scalar field between two parallel conducting plates. Next we evaluate the Casimir energy by summing over the modes for the massless (Sec. \ref{CasimirMasslessSec}) and the massive (Sec. \ref{CasimirMassiveSec}) cases in Sec. \ref{CasimirEffect}. As a nature of validation, we show that the expressions for the Casimir energy reduce to the well-known results in the conventional scalar field. Details of technical computations are relegated to the Appendix. Finally in Sec. \ref{ConclusionSection}, we summarize the main results of the paper and discuss how they compare with those obtained in alternative theories for quantum gravity, such as GUPs.

\section{Polymer quantization} \label{PolScalarField}

As noted above, the central difference between Schr\"{o}dinger and polymer quantization is the choice of Hilbert space. In polymer representation, the corresponding Hilbert space $\mathcal{H} _{\mbox{\scriptsize poly}}$ is the Cauchy completion of the set of linear combination of some basis states $ \left\lbrace \left\vert x_{\mu }\right\rangle \right\rbrace  $, with inner product
\begin{equation}
\left\langle x_{\mu} \vert x_{\nu} \right\rangle = \lim _{T \rightarrow \infty} \frac{1}{2T} 
\int _{-T} ^{T} e^{-i \frac{p}{\hbar} \left( x_{\mu} - x_{\nu} \right)} \, dp = \delta _{\mu 
\nu}, \label{inner}
\end{equation}
where the right-hand side is the generalization of the Kronecker delta to an uncountable index set. Plane waves are normalizable in this inner product. The kinematical Hilbert space can be written as $ \mathcal{H} _{\mbox{\scriptsize poly}} = L^{2} \left(  \mathbb{R} _{d}, d \mu _{d} \right) $, with $  d \mu _{d} $ corresponding to the Haar measure, and $ \mathbb{R} _{d} $ the real line endowed with the discrete topology.

The state of a polymer system can be expressed as
\begin{equation}
\left\vert \psi \right\rangle = \sum _{\mu = - \infty} ^{+ \infty} \psi _{\mu}  \left\vert x_{\mu 
}\right\rangle ,
\end{equation}
where $ \left\vert x_{\mu }\right\rangle $ are eigenstates of the position operator
\begin{equation}
x \left\vert x_{\mu }\right\rangle = x_{\mu } \left\vert x_{\mu }\right\rangle ,
\end{equation}
and the $\psi _{\mu}$'s are expansion coefficients. Note that the spectrum of the position operator $ \left\lbrace x _{\mu} \right\rbrace  $ consists of a countable selection of points from the real line $ \mathbb{R} $, which is analogous to the graph covering $3$-manifolds in LQG.

The central feature here is that the momentum operator $p$ is not realized directly as in Schr\"{o}dinger representation because of built-in notion of discreteness, but arise indirectly through translation operator $U _{\lambda} = e ^{i \frac{p \lambda}{\hbar}} $. Here, $\lambda$ is a fundamental length scale of the theory. Hence, for the representation of the Heisenberg-Weyl algebra we choose the position operator $x$ and the translation operator $U _{\lambda}$ instead of the momentum operator. The action of the translation operator on position eigenstates is
\begin{equation}
U _{\lambda}  \left\vert x_{\mu }\right\rangle = \left\vert x_{\mu } - \lambda 
\right\rangle ;
\end{equation}
that is, $U _{\lambda} $ converts a position eigenstate with eigenvalue $ x_{\mu} $ into an eigenstate with eigenvalue $ x _{\mu} -\lambda $. These operators definitions give the basic commutator $ [x , U  _{\lambda} ] = - \lambda  U _{\lambda} $, and $U _{\lambda} $ defines a one-parameter family of unitary operators on $\mathcal{H} _{\mbox{\scriptsize poly}}$, where its adjoint is given by $U ^{\dagger} _{\lambda} = U _{- \lambda}  $. Mathematically, polymer and Schr\"{o}dinger quantizations are inequivalent because $U _{\lambda} $ is discontinuous with respect to $ \lambda $ given that $ \left\vert x_{\mu }\right\rangle $ and $ \left\vert x_{\mu } - \lambda \right\rangle $ are always orthogonal, no matter how small is $ \lambda $ \cite{Corichi}.

However, inspired by the techniques used in Lattice Gauge Theories and LQG, by introducing a fixed length scale $ \lambda $ it is possible to define an effective momentum operator as follows
\begin{equation}
p _{\lambda} = \frac{\hbar}{2 i \lambda} \left( U _{\lambda} - U ^{\dagger} _{\lambda} \right) , \label{momentumOp}
\end{equation}
which corresponds to the approximation $ p \lambda \ll \hbar $. Practical calculations in polymer quantum mechanics involve the mapping $p \to p _{\lambda}$ in any momentum-dependent operator. In $L ^{2} (\mathbb{R},dx)$, the $\lambda \to 0 $ limit would give the usual momentum $- i \hbar \partial _{x}$ and momentum-squared $- \hbar ^{2} \partial _{x} ^{2}$ operators. In $ \mathcal{H} _{\mbox{\scriptsize poly}} = L^{2} \left(  \mathbb{R} _{d}, d \mu _{d} \right) $ this limit does not exist because $\lambda$ is regarded as a fundamental length scale. This is analogous to the quantum-classical transition through $\hbar \to 0 $ limit, where $\hbar$ is a nonzero fundamental constant of quantum theory \cite{AMR7, AMR8}.

Regularizing the kinetic energy operator $K =p ^{2} / 2m$, the polymer representation of the classical non-relativistic Hamiltonian $H = K+V$ is then \cite{Ashtekar} 
\begin{equation}
H _{\lambda} = \frac{\hbar ^{2}}{2m \lambda ^{2}} \left( 2 - 
U _{\lambda} - U ^{\dagger} _{\lambda} \right)  + 
V \left( x \right) , \label{hamiltonian}
\end{equation}
where the potential term is arbitrary but assumed to be regular so that $ V$ can be defined by pointwise multiplication, $ \left\langle x_{\mu} \left\vert V \right\vert \psi \right\rangle = V \left( x _{\mu} \right) \left\langle x_{\mu} \vert \psi \right\rangle  $.

The dynamics of the system will be determined by the equation $i \hbar \partial _{t} \left\vert \psi \right\rangle = H _{\lambda} \left\vert \psi \right\rangle$, which decomposes the polymer Hilbert space $\mathcal{H} _{\mbox{\scriptsize poly}}$, into an infinite superselected finite-dimensional subspaces, each with support on a regular lattice $ \gamma = \gamma \left( \lambda , x _{0} \right)  $ with the same space between points $ \lambda$, where $ \gamma \left( \lambda , x _{0} \right) = \left\lbrace n \lambda + x_{0} \vert n \in\mathbb{Z} \right\rbrace  $, and $ x _{0} \in \left[ 0 , \lambda \right)  $. This way of choosing $ x _{0} $ fixes the superselected sector, restricting the dynamics to a lattice $ \gamma \left( \lambda , x _{0} \right)   $ and work on separable Hilbert space $\mathcal{H} _{\mbox{\scriptsize poly}} ^{x _{0}}$ consisting of wave functions which are nonzero only on the lattice.

The energy eigenvalue equation $H _{\lambda} \left\vert \psi \right\rangle = E _{\lambda} \left\vert \psi \right\rangle$ becomes a difference equation for the coefficients $\psi _{\mu} = \left\langle x _{\mu} \vert \psi \right\rangle$ in the coordinate representation:
\begin{equation}
\psi _{\mu + 1} + \psi _{\mu - 1} = 2 \left\lbrace  1 -\frac{ m \lambda ^{2}}{\hbar ^{2}} \left[ E _{\lambda} - V \left( x _{\mu} \right)  \right] \right\rbrace  \psi _{\mu}.  
\end{equation}
On the other hand, in the momentum representation, it is generically a differential equation for $ \psi \left( p \right) = \left\langle p \vert \psi \right\rangle$ :
\begin{equation}
\frac{\hbar ^{2}}{m \lambda ^{2} } \left[  1- \cos \left( \frac{p \lambda}{\hbar} \right) \right] \psi \left( p \right) = \left[ E _{\lambda} - V \left( -i \hbar \partial _{p} \right) \right] \psi \left( p \right) ,
\end{equation}
where we have used that $\left\langle p \right\vert \! x \! \left\vert \psi \right\rangle = i \hbar \partial _{p} \psi \left( p \right)$ and $\left\langle p \right\vert \! U _{\lambda} \! \left\vert \psi \right\rangle =e ^{i p \lambda / \hbar} \; \psi \left( p \right)$. 

Working on $ \gamma \left( \lambda , x _{0} \right) $ restricts momentum wave functions $ \psi \left( p \right) $ to periodic functions of period $2 \pi \hbar / \lambda$ with the inner product formula (\ref{inner}) reducing to:
\begin{equation}
\left\langle x_{\mu} \vert x_{\nu} \right\rangle = \left\langle x_{\mu } \right\vert \Bigg( 
\frac{\lambda}{2 \pi  \hbar} \int _{- \pi \hbar / \lambda} ^{\pi \hbar / \lambda} \! \vert p  \left\rangle  \right\langle  p \vert \, dp \Bigg) \left\vert 
x_{\nu}\right\rangle = \delta _{\mu \nu} , \label{inner2}
\end{equation}
and $ p \in \left( - \pi \hbar / \lambda , \pi \hbar / \lambda \right)$. Note that the identity operator (readed from (\ref{inner2})) on such subspace serves to define the inner product on $\mathcal{H} _{\mbox{\scriptsize poly}} ^{x _{0}}$ in the momentum representation.

There are several examples of systems that have been treated with polymer quantization. These are simple quantum mechanical systems, such as the harmonic oscillator \cite{Ashtekar2}, the particle in a box \cite{Corichi}, the diffraction in time \cite{AMR1}, the tunneling through potential barriers \cite{AMR2}, the Coulomb potential \cite{Husain, Kunstatter} and the quantum bouncer \cite{AMR3}. Using these results, the authors in Ref. \cite{Chacon} have studied the statistical thermodynamics of such systems.

PQ has also been used to study a simple quantum cosmological model known as polymer quantum cosmology \cite{Corichi}. The main idea behind this model is that the phase space of cosmological spacetimes that are homogeneous and isotropic (and for which the homogeneous spatial slices have flat intrinsic geometry, i.e. the Friedmann-Robertson-Walker cosmology), can be polymer-quantized following the theory described above. After some appropriate considerations and with the correct choice of polarization, this yield to a polymer Wheeler-DeWitt equation  \cite{Corichi}.

\section{Confined polymer scalar field} \label{ConfinedPolScalField}

We now turn to the study of the three-dimensional Klein-Gordon equation in polymer representation. To this end, we restrict our analysis to an orthorhombic primitive Bravais lattice $\gamma (\vec{\lambda} , \vec{r} _{0}) = \left\lbrace \vec{r} _{0} + q \lambda _{x} \vec{e} _{x} + t \lambda _{y} \vec{e} _{y} + n \lambda _{z} \vec{e} _{z} \;  \vert \; (q,t,n) \in \mathbb{Z} \right\rbrace$, where $\lambda _{x}$, $\lambda _{y}$ and $\lambda _{z}$ are the three independent lattice parameters \cite{Kittel}, and $\vec{r} _{0}$ is a vector in the primitive cell (i.e. $\vert \vec{r} _{0} \vert < \mbox{min} (\lambda _{x} , \lambda _{y} , \lambda _{z})$) which fixes the superselected sector. For the sake of simplicity here we take $\vec{r} _{0} = \vec{0}$, although the following analysis is also valid for an arbitrary vector $\vec{r} _{0}$.

Just as we did to derive the polymer representation of the Schr\"{o}dinger (\ref{hamiltonian}) and the Wheeler-DeWitt equations \cite{Corichi}, the regularization of the classical Klein-Gordon Hamitonian $H ^{2} = (\vec{p} \, c) ^{2} + (mc ^{2}) ^{2}$ produces
\begin{align}
H ^{2} _{\vec{\lambda}} = \hbar ^{2} c ^{2} \sum _{i=x,y,z} \frac{1}{\lambda ^{2} _{i}} \left( 2 - 
U _{\lambda _{i}} - U ^{\dagger} _{\lambda _{i}} \right) + (mc ^{2}) ^{2} , \label{PolymerKG}
\end{align}
where $U _{\lambda _{i}} = e ^{i p _{i} \lambda _{i} / \hbar}$ is the translation operator along the $\vec{e} _{i}$-direction. The problem now consists in solving the eigenvalue equation $H _{\vec{\lambda}} ^{2} \left\vert \psi \right\rangle = E _{\vec{\lambda}} ^{2} \left\vert \psi \right\rangle$ for our Casimir system, which consists in two parallel conducting plates separated by a distance $L$ along a given direction, say the $\vec{e} _{z}$-direction. The minimum length scale in the $\vec{e} _{z}$-direction requires $L = N \lambda _{z}$, with $N \in \mathbb{Z}$.

We assume that the scalar field satisfies Dirichlet boundary conditions on the plates, located at $z = 0$ and $z = L$, having lattice positions $n = 0$ and $n = N $. Due to the translational invariance in the directions parallel to the plates, that is in the transverse $x$ and $y$ directions, we can work in a mixed coordinate-momentum representation $\psi _{n} (p _{x} , p _{y}) = \left\langle p _{x} , p _{y} , z _{n} \vert \psi \right\rangle$. This allow us to write the following difference equation for the function $\psi _{n}$:
\begin{align}
\mathcal{E} _{\vec{\lambda}} ^{2} \, \psi _{n} = \frac{\hbar ^{2} c ^{2}}{\lambda _{z} ^{2}} \left( 2 \psi _{n} - \psi _{n+1} - \psi _{n-1}  \right) , \label{DiffEq}
\end{align}
where we have defined
\begin{align}
\mathcal{E} _{\vec{\lambda}} ^{2} = E _{\vec{\lambda}} ^{2} \! - \! (mc ^{2}) ^{2} \! - \! \hbar ^{2} c ^{2} \! \left( \frac{4}{\lambda _{x} ^{2}} \sin ^{2} \frac{p _{x} \lambda _{x}}{2 \hbar} +  \frac{4}{\lambda _{y} ^{2}} \sin ^{2} \frac{p _{y} \lambda _{y}}{2 \hbar} \right) \! , \label{E-Cal}
\end{align}
with $ p _{x} \in \left( - \pi \hbar / \lambda _{x} , \pi \hbar / \lambda _{x} \right)$ and $ p _{y} \in \left( - \pi \hbar / \lambda _{y} , \pi \hbar / \lambda _{y} \right)$.

Now we have to solve the second order difference equation (\ref{DiffEq}) subject to the boundary conditions $\psi _{n = 0} = 0$ and $ \psi _{n = N} = 0$. To this end, and following \cite{Elaydi}, we propose the solution of Eq. (\ref{DiffEq}) to be 
\begin{align}
\psi _{n} = \alpha _{1} \, r ^{n} _{1} + \alpha _{2} \, r ^{n} _{2} , \label{solution}
\end{align}
where $\alpha _{i}$ are constant coefficients and $r _{i}$ are the roots of the characteristic equation:
\begin{align}
r ^{2} - \left( 2 - \kappa ^{2} \right) r + 1 = 0 ,
\end{align}
whose solutions are
\begin{align}
r _{\pm} = 1 - ( \kappa ^{2} / 2) \pm (1/2) \sqrt{ \kappa ^{2} \left( \kappa ^{2} - 4 \right)} , \label{Roots}
\end{align}
where $\kappa = \mathcal{E} _{\vec{\lambda}} \lambda _{z} / \hbar c$. This expression gives us a relation between energy $\mathcal{E} _{\vec{\lambda}}$ and $2 \hbar c / \lambda _{z}$. If $\vert \mathcal{E} _{\vec{\lambda}} \vert > 2 \hbar c / \lambda _{z}$, then the roots $r _{\pm}$ are real numbers (including the degenerate case), but incompatible with the boundary conditions, therefore they yield the trivial solution $\psi _{n} = 0$. This result implies that the minimum length imposes a cutoff on the energy, and hence, energies beyond the maximum are unphysical. The only physical solutions are those for which $\vert \mathcal{E} _{\vec{\lambda}} \vert < 2 \hbar c / \lambda _{z}$, in which case the roots (\ref{Roots}) are complex. In this case, the solution can be written as
\begin{align}
\psi _{n} = \beta _{1} \, \cos (n \theta) + \beta _{2} \, \sin (n \theta) , \label{solution2}
\end{align}
which is a parametrization of (\ref{solution}) in polar coordinates, with
\begin{align}
\cos \theta = \frac{1}{\sqrt{2}} \frac{2 - \kappa ^{2}}{\sqrt{\kappa ^{4} - 4 \kappa ^{2} + 2}} .
\end{align}
Imposing the boundary conditions $\psi _{n = 0} = 0$ and $\psi _{n=N} = 0$ in (\ref{solution2}) yields the condition $N \theta = s \pi$, with $s \in \mathbb{Z} $. The solution to the difference equation (\ref{DiffEq}) turns out to be
\begin{align}
\psi _{n,s} = \mathcal{N} \sin \left( n \pi \frac{s}{N} \right) , \qquad 0 < s < N ,  \label{eigenfunction}
\end{align}
where $\mathcal{N}$ is a normalization factor
\begin{align}
\mathcal{N} = \left[ \sum _{s = 0} ^{N} \sin ^{2} \left( \frac{n \pi s}{N}  \right) \right] ^{-1/2} = \sqrt{\frac{2}{N}} .
\end{align}
The corresponding energy spectrum (which can be obtained by substituting the eigenfunction (\ref{eigenfunction}) into the difference equation (\ref{DiffEq})) is found to be bounded:
\begin{align}
 \mathcal{E} _{\vec{\lambda} , s} =  \frac{2 \hbar c }{\lambda _{z}}  \sin \left( \frac{\pi s}{2 N} \right) ,
\end{align}
which resembles the tight binding model of a particle in a periodic potential with periodic boundary conditions \cite{Kittel}. Inserting this result in Eq. (\ref{E-Cal}) we obtain the full energy spectrum for a polymer scalar field confined between two parallel conducting plates:
\begin{align}
E _{\vec{\lambda} , s} &= \Bigg[ (mc ^{2}) ^{2} + \hbar ^{2} c ^{2} \! \left( \frac{4}{\lambda _{x} ^{2}} \sin ^{2} \frac{p _{x} \lambda _{x}}{2 \hbar} +  \frac{4}{\lambda _{y} ^{2}} \sin ^{2} \frac{p _{y} \lambda _{y}}{2 \hbar} \right. \notag \\  & \hspace{3.5cm} \left. + \frac{4}{\lambda _{z} ^{2}} \sin ^{2} \frac{\pi s}{2 N}  \right) \Bigg] ^{1/2} . \label{PolimericEnergy}
\end{align}

\section{The Casimir effect} \label{CasimirEffect}

In this section we will compute the Casimir energy and stress (per unit area) associated with a polymer scalar field confined between two parallel conducting plates separated by a distance $L$ along the $z$ direction. In general, the lattice parameters $\lambda _{i}$ can be different, specially in solid-state systems. Nevertheless, from a quantum gravity point of view, there is no evidence of a preferred spatial direction, and hence we can safely take $\lambda _{x} = \lambda _{y} = \lambda _{z} \equiv \lambda$. 

There are many ways in which the Casimir effect can be computed. Perhaps the most obvious procedure is to compute the zero-point energy in the presence of the plates. This is precisely the method we shall employ in this paper. On the one hand, we will consider the polymer scalar field interacting with the plates at $z = 0$ and $z = L$. On the other hand, we must consider contributions from the field inside and outside the plates as well \cite{Gambini}. To do this, we introduce two auxiliary plates at $z = L _{1}$ and $z = - L _{2}$, such that $L _{1} \gg L$. In the following, we consider the massless and massive cases separately.

\subsection{Massless case} \label{CasimirMasslessSec}
 
The zero-point energy of a massless polymer scalar field inside the cavity will be given according to (\ref{PolimericEnergy}) by
\begin{align}
\mathcal{E} (L) = \frac{\hbar c}{\lambda}  \int _{\mathcal{D}} \frac{d ^{2} \vec{k}}{(2 \pi) ^{2}} \sum _{n = 1} ^{N-1} \Bigg[  \sin ^{2}  \frac{k _{x} \lambda}{2} + \sin ^{2} \frac{k _{y} \lambda}{2}   \notag \\ + \sin ^{2} \frac{n \pi}{2 N} \Bigg] ^{1/2}  , \label{0EnergyMassless}
\end{align}
where $ \vec{k} = k _{x} \vec{e} _{x} + k _{y} \vec{e} _{y} = \vec{p} / \hbar$ is the wave-vector parallel to the plates and $\mathcal{D}$ is the rectangular domain defined as
\begin{align}
\mathcal{D} \equiv \left\lbrace (k _{x} , k _{y} ) \in \mathbb{R} ^{2} : - \frac{\pi}{\lambda} \leq k _{x} \leq  \frac{\pi}{\lambda} ; - \frac{\pi}{\lambda} \leq k _{y} \leq  \frac{\pi}{\lambda} \right\rbrace . \label{RegionD}
\end{align}
We observe that in the limit $\lambda / L \to 0$, Eq. (\ref{0EnergyMassless}) correctly reduces to the zero-point energy of a massless scalar field \cite{Milton, Bordag} and the rectangular domain $\mathcal{D}$ opens to the whole $\mathbb{R} ^{2}$ \cite{AMR1, Corichi}, as it should be.

In order to evaluate the integrals appearing in Eq. (\ref{0EnergyMassless}), it is convenient to employ the Schwinger proper-time representation for the square root \cite{Milton}:
\begin{align}
\mathcal{E} (L) &=  \frac{\hbar c}{2 \Gamma(-\frac{1}{2})} \int _{\mathcal{D}} \frac{d ^{2} \vec{k}}{(2 \pi) ^{2}} \sum _{n = 1} ^{N-1} \int _{0} ^{\infty}  dx \;  x ^{-3/2} \notag \\ & \phantom{=}  \times \! \exp{ \! \left\lbrace  \! - \frac{4 x}{\lambda ^{2}}  \left( \sin ^{2} \! \frac{k _{x} \lambda}{2} + \sin ^{2} \! \frac{k _{y} \lambda}{2} +\sin ^{2} \! \frac{n \pi}{2 N} \right) \! \right\rbrace }  \label{0EnergyMassless2}
\end{align}
where we have used the Euler representation for the gamma function. We next carry out the integration over the momentum $\vec{k} $. The result is simple and with a change of variables it can be cast into the form
\begin{align}
 \mathcal{E} (L) &= \frac{\hbar c}{2 L ^{3} \Gamma(-\frac{1}{2})} \sum _{n = 1} ^{N-1} \int _{0} ^{\infty} dx \;  x ^{-3/2} e ^{- 4 x \left( N \sin  \frac{n \pi}{2 N} \right) ^{2} } \notag \\ & \hspace{2.5cm} \times \left[ N \,  I _{0} \left( 2xN ^{2} \right) e ^{- 2 xN ^{2}} \right] ^{2} , \label{0EnergyMassless3}
\end{align}
where $I _{0}(x)$ is the zeroth-order modified Bessel function of the first kind and $N = L / \lambda$ is the number of points between the plates. The integral and summation in Eq. (\ref{0EnergyMassless3}) cannot be calculated explicitly, therefore we will consider their asymptotic limits. This can be done safely since up to date there is not experimental evidence of a fundamental length scale, and therefore we should assume $\lambda$ to be small as compared with any other relevant physical length in the problem (the separation $L$ between the plates, for instance). Let us proceed along this way. Taking the asymptotic behavior of the integrand in Eq. (\ref{0EnergyMassless3}) for $N \gg 1$ and keeping terms up to order $1 / N ^{2}$ we obtain
\begin{align}
\mathcal{E} (L) & \approx  \frac{\hbar c}{2 L ^{3} \Gamma(-\frac{1}{2})} \sum _{n = 1} ^{N-1} \int _{0} ^{\infty}  dx \;  x ^{-3/2} e ^{- x n ^{2} \pi ^{2}} \notag \\ & \hspace{0.7cm} \times \frac{1}{4 \pi x} \left( 1 + x \frac{n ^{4} \pi ^{4}}{12 N ^{2}} \right) \left( 1 + \frac{1}{8 x N ^{2}} \right) , \label{0EnergyMassless4}
\end{align}
where we have used that $I _{0} (z) \sim \frac{e ^{z}}{\sqrt{2 \pi z}} \left( 1 + \frac{1}{8 z} \right) $ for $z \gg 1$ and $\sin z \approx z - \frac{z ^{3}}{3!}$ for $z \ll 1$ \cite{Ryzhik}. For the sake of simplicity, we write the energy per unit area in the simplest form
\begin{align}
\mathcal{E} (L) & = - \frac{\pi ^{2}}{1440} \frac{\hbar c}{L ^{3}} \left( A _{1} + \frac{A _{2}  + A _{3} }{N ^{2}} \right) , \label{0EnergyMassless5}
\end{align}
where
\begin{align}
A _{1} &= - \frac{180}{\pi ^{3} \Gamma (-\frac{1}{2})} \sum _{n = 1} ^{N-1} \int _{0} ^{\infty} dx \;  x ^{-5/2} e ^{- x n ^{2} \pi ^{2}} , \label{A1} \\ A _{2} &= - \frac{15 \pi}{ \Gamma (-\frac{1}{2})} \sum _{n = 1} ^{N-1} n ^{4} \int _{0} ^{\infty} dx \;  x ^{-3/2} e ^{- x n ^{2} \pi ^{2}}  , \label{A2} \\ A _{3} &= - \frac{45}{2 \pi ^{3} \Gamma (-\frac{1}{2})} \sum _{n = 1} ^{N-1} \int _{0} ^{\infty} dx \;  x ^{-7/2} e ^{- x n ^{2} \pi ^{2}}  . \label{A3}
\end{align}
These expressions can be further simplified. Making use of the following integral representation of the gamma function \cite{Ryzhik}
\begin{align}
\int _{0} ^{\infty} x ^{s-1} e ^{- \alpha x} dx = \frac{\Gamma (s)}{\alpha ^{s}} , \label{Exp-Int}
\end{align}
we get
\begin{align}
A _{1} = 120 \sum _{n = 1} ^{N-1}  n ^{3} , \quad A _{2} = \frac{15}{6} A _{3} = - 15 \pi ^{2}  \sum _{n = 1} ^{N-1}  n ^{5} . \label{A123}
\end{align}
The problem has been reduced now to the evaluation of these partial sums. Before embarking us in such technical problem, let us verify the continuous limit. To this end, we take the number of points between the plates very large but keeping fixed its distance, i.e. $N \to \infty$ and $N \lambda \to L$. In this case the upper limit in the summations of Eq. (\ref{A123}) opens to infinity and the apparently non convergent sums can be resolved by means of the analytic continuation of the Riemann zeta function $\zeta (s)$. By direct calculation we find that, in the continuous limit, $A _{1} = 120 \, \zeta (-3) = 1$, $A _{2} = -15 \pi ^{2} \, \zeta (-5) = 5 \pi ^{2}/84$ and $A _{3} = - 6 \pi ^{2} \, \zeta (-5) = \pi ^{2} / 42$. Inserting these results into Eq. (\ref{0EnergyMassless5}) we can see that only the first term survives to the limit $N \to \infty$ (due to the $1/N$ expansion), which implies that the energy per unit area (\ref{0EnergyMassless5}) correctly reduces to the continuous result for a massless scalar field \cite{Milton}. 

Now let's go back to the evaluation of the partial sums in Eq. (\ref{A123}). To this end, we begin by discussing how the definition of the Riemann zeta function, 
\begin{align}
\zeta (s) = \sum _{n = 1} ^{\infty} \frac{1}{n ^{s}} , \label{RiemannZ}
\end{align}
which is an absolutely convergent serie in the region $\mbox{Re} (s) > 0$, can be extended to the case $\mbox{Re} (s) < 0$ by purely real-variable methods. Of course, we know in advance that the $\zeta$ function can be extended to this region (with a pole at $s=1$) by analytic continuation; nevertheless real-variable methods allow us to approximate the partial sums appearing in Eq. (\ref{A123}) by a polynomial in $1/N$. See Ref. \cite{Terence} for a detailed discussion.

To put this problem on the table, let's notice that the summations in Eq. (\ref{A123}) do not make sense if we stay within the traditional way to evaluate infinite series. For example, it is well known that $\sum _{n = 1} ^{N-1} \! n ^{3}  = \frac{1}{4} N ^{2} (N-1) ^{2}$ and $\sum _{n = 1} ^{N-1} \! n ^{5}  = \frac{1}{12} N ^{2} (N-1) ^{2} (2N ^{2}-2N-1)$. Accordingly the limit $N \to \infty$ produces infinities, which are indeed inconsistent with the finite results obtained via the analytic continuation of the Riemann $\zeta$ function (\ref{RiemannZ}). This problem can be resolved with real-variable methods by replacing the abruptly truncated sums $\sum _{n = 1} ^{N-1}n ^{s}$ with smoothed sums $\sum _{n = 1} ^{\infty} n ^{s} \eta (n/N)$, where $\eta (x): \mathbb{R} ^{+} \to \mathbb{R}$ is a twice continuously differentiable cutoff function satisfying $\eta (x) \to 1$ pointwise as $x \to 0$ (a condition required to fulfill the limit $N \to \infty$) and is uniformly bounded \cite{Terence}. This program produces an expansion of the sum in powers of $1/N$, which is treated as a small parameter. Performing a $1/N$ expansion gets more and more accurate in the large $N$ limit. Remarkably, the dominated convergence theorem guarantees that smoothing does not affect the asymptotic value of the sum (i.e. the zeroth order term of the expansion), which is absolutely convergent. Higher order terms are to be understood as correction terms. In general, smoothing is a conceptual bridge between zeta function regularization, with its reliance on complex analysis, and Ramanujan summation, with its shortcut to the Euler-Maclaurin formula \cite{Terence}. 

The most famous smoothed sum is perhaps the Ces\`{a}ro summation, which corresponds to the cutoff function $\eta _{\mbox{\tiny C}} (x) = (1 -x) _{+}$. This correctly gives the well-known result $1/2$ of the Grandi’s series. In physics, an exponential cutoff is often much more useful since it decays more rapidly than the Ces\`{a}ro sum. In fact, the Ces\`{a}ro cutoff function corresponds to the leading term of the exponential cutoff $\eta _{\mbox{\tiny E}} (x) = e ^{-x}$ when it is Taylor expanded.  In this paper we shall use the exponential cutoff function to evaluate the partial sums of Eq. (\ref{A123}). Even more, since we have computed the Casimir energy up to order $1 / N ^{2}$, we will Taylor expand the cutoff function $\eta _{\mbox{\tiny E}} (n/N)$ up to order $1 / N ^{2}$ in the following approximations. Applying the above discussed program we obtain
\begin{align}
\sum _{n = 1} ^{N-1} \! n ^{s} \! &= \! \sum _{n = 1} ^{\infty} \!  n ^{s} \, \eta _{\mbox{\tiny E}} ( n /N ) \! \approx \! \sum _{n = 1} ^{\infty} \! n ^{s} \! \left[1 \! - \! \frac{n}{N} \! + \! \frac{n ^{2}}{2 N ^{2}} \! + \! \mathcal{O} (N ^{-3}) \right] \notag \\ &= \zeta (-s) - \frac{\zeta (-s-1)}{N} + \frac{\zeta (-s-2)}{2 N ^{2}}  +  \mathcal{O} ( N ^{-3}) ,
\end{align}
where $\zeta (-s)$ is the Riemann $\zeta$ function (\ref{RiemannZ}) analytically continued to negative arguments. This result implies that the coefficients (\ref{A123}) become
 \begin{align}
A _{1} = 1 - \frac{5}{21 N ^{2}} , \;\; A _{2} = \frac{15}{6} A _{3} = \frac{\pi ^{2}}{672} \left( 40 - \frac{21}{N ^{2}} \right) , \label{A123-2}
\end{align}
where we have used that $\zeta (-2s) = 0$ for $s \in \mathbb{Z} ^{+}$. Substituting these results into Eq. (\ref{0EnergyMassless5}) and keeping terms up to order $1/N ^{2}$ we obtain
\begin{align}
\mathcal{E} (L) & = - \frac{\pi ^{2}}{1440} \frac{\hbar c}{L ^{3}} \left[ 1 + \frac{1}{N ^{2}} \left( \frac{ \pi ^{2}}{12}  - \frac{5}{21} \right) \right] . \label{0EnergyMasslessFIN}
\end{align} 
The first term in Eq. (\ref{0EnergyMasslessFIN}) is the usual Casimir energy for a massless scalar field, while the second term is the correction given by the presence of the minimal length of the polymer theory. We note that it is attractive. 

\begin{figure}
\includegraphics[scale=0.45]{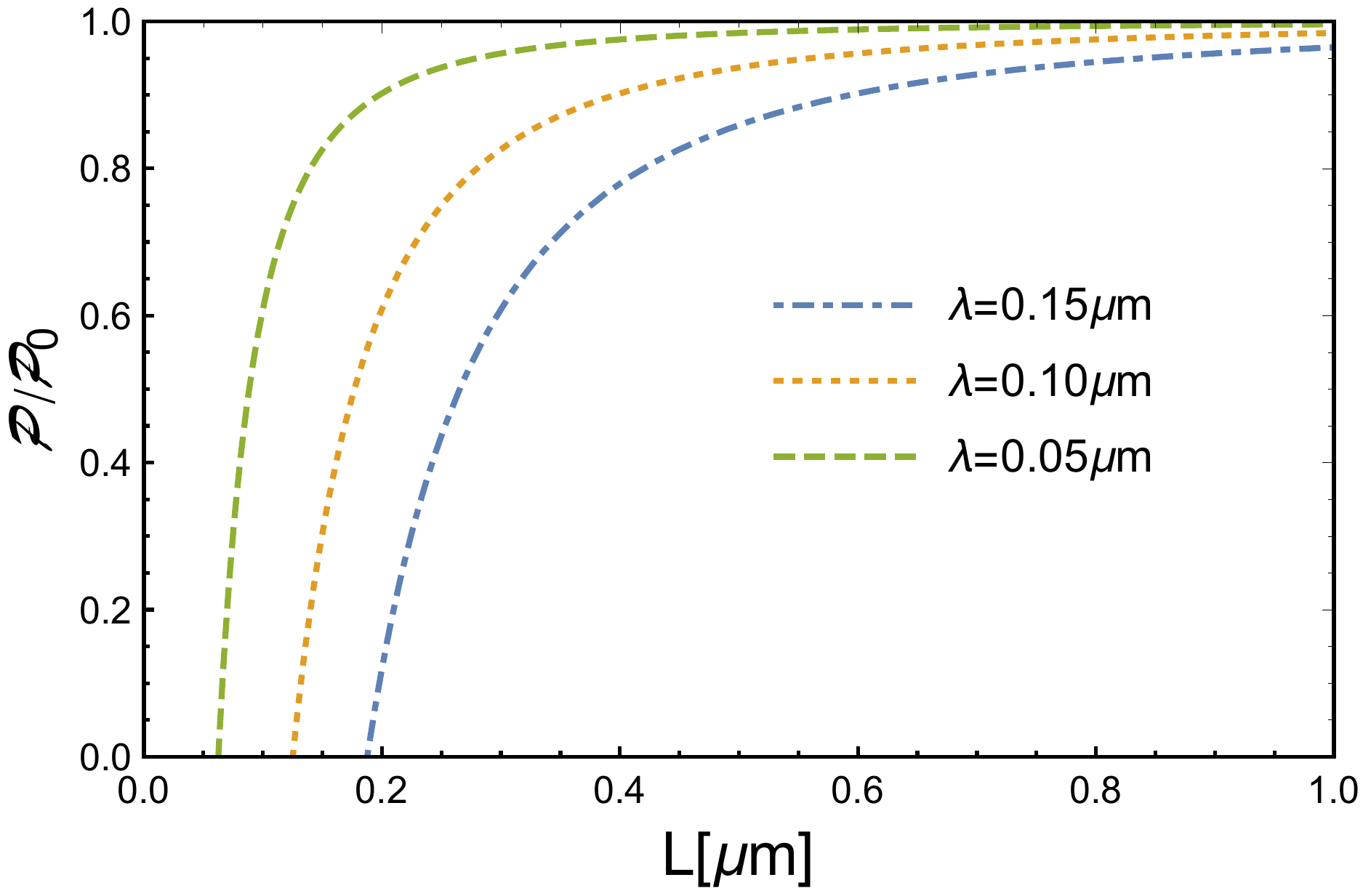}
\caption{Casimir pressure for a massless polymer scalar field (in units of the usual Casimir pressure $\mathcal{P} _{0}$) as function of $L$ for different values of the minimal length of the polymer theory.}
\label{CasForceMassless}
\end{figure}

Now, using the result of  Eq. (\ref{0EnergyMasslessFIN}) we can obtain the total energy of the system. For the scalar field inside the region between the plate at $z = L$ and the auxiliary plate at $z= L _{1}$ the same construction can be adopted, and the energy density is given by Eq. (\ref{0EnergyMasslessFIN}) with the replacement $L \to L _{1} - L$. For the slab of width $L _{2}$, between the plate at $z=0$ and the auxiliary plate at $z = - L _{2}$, the energy density is also given by Eq. (\ref{0EnergyMasslessFIN}) with $L \to L _{2}$. The total energy of the system will be then
\begin{align}
\mathcal{E} _{\mbox{\scriptsize T}} (L) = \mathcal{E} (L) + \mathcal{E} (L _{1} - L)  + \mathcal{E} (L _{2}) .   \label{TotalEnergy}
\end{align}
The total Casimir pressure upon the plate at $z=L $ is given by $\mathcal{P} _{\mbox{\scriptsize T}} (L) = - \frac{1}{2 \lambda} \left[ \mathcal{E} _{\mbox{\scriptsize T}} (L+ \lambda) - \mathcal{E} _{\mbox{\scriptsize T}} (L - \lambda) \right]$. Since we are interested in variations with respect to $L$, the last term in Eq. (\ref{TotalEnergy}) does not contribute to the pressure. Keeping terms up to order $1/N ^{2}$ we obtain
\begin{align}
\mathcal{P} _{\mbox{\scriptsize T}} (L) = \mathcal{P} (L) + \mathcal{P} (L _{1} - L) , \label{TotalPressure}
\end{align}
where
\begin{align}
\mathcal{P} (L) = - \frac{\pi ^{2}}{480} \frac{\hbar c}{L ^{4}} \left[ 1 + \frac{5 \lambda ^{2}}{9 L ^{2}} \left( \frac{\pi ^{2}}{4} - \frac{37}{7} \right) \right] . \label{ForceMasslessFIN}
\end{align}
For a finite value of $L _{1}$, Eq. (\ref{TotalPressure}) gives the pressure upon the plate at $z = L$ when placed between the plates at $z = 0$ and $z = L _{1}$, i.e. we have a Casimir piston. In order to obtain the Casimir force for our initial configuration we take the limit $L _{1} \gg L$, such that $\mathcal{P} _{\mbox{\scriptsize T}} (L) = \mathcal{P} (L)$. In Fig. \ref{CasForceMassless} we plot the Casimir pressure $\mathcal{P}$ (in units of $\mathcal{P} _{0} = - \frac{\pi ^{2}}{480} \frac{\hbar c}{L ^{4}}$, which is the usual attractive pressure for a massless scalar field) as a function of the plate's separation $L$ for different values of $\lambda$ (here, both $L$ and $\lambda$ are measured in the same units of length). As we can see, the Casimir pressure for a massless scalar polymer field tends asymptotically to the usual Casimir pressure as $\lambda$ approaches to zero, as expected.

We can make some considerations about the possibility of observing this effect. Clearly, if $\lambda$ is of the order of the Planck length $\sim 10 ^{-35}$m (although no known data substantiate this conjecture), no observation is possible. However, current experiments on the Casimir force can set an upper bound on the minimal length of the polymer theory. The authors in Ref. \cite{Bressi} measured the Casimir force between parallel plates, with the distance $L$ between the surfaces in the range $0.5-3 \mu$m and a precision of $15 \%$. Using this result, the upper bound obtained for the minimal length scale is $\lambda = 1.5 \times 10 ^{-7}$m, which is far from the one expected if we think $\lambda$ as arising from a more fundamental physical theory. Indeed, this bound is comparable with the one obtained by using the diffraction in time \cite{AMR1}, the polymer bouncer \cite{AMR3} and by means of polymer Bose-Einstein condensate \cite{Chacon2}.

\subsection{Massive case} \label{CasimirMassiveSec}

It is simple but not straightforward to extend the discussion of Sec. \ref{CasimirMasslessSec} to include the mass $m$ for the scalar field. The zero-point energy inside the cavity now becomes
\begin{align}
\mathcal{E} (m, L) &= \frac{c}{2}  \int _{\mathcal{D}} \frac{d ^{2} \vec{k}}{(2 \pi) ^{2}} \sum _{n = 1} ^{N-1}  \Bigg[ \left( m c \right) ^{2} + \left(  \frac{2 \hbar}{\lambda} \right) ^{2} \notag \\ & \phantom{=} \times \left( \sin ^{2} \! \frac{k _{x} \lambda}{2} + \sin ^{2} \! \frac{k _{y} \lambda}{2} +\sin ^{2} \! \frac{n \pi}{2 N} \right) \Bigg] ^{1/2}   , \label{0EnergyMassive}
\end{align}
where $\mathcal{D}$ is the rectangular domain defined in Eq. (\ref{RegionD}). The calculation proceeds just as in the massless case: we first introduce the Schwinger proper time representation for the square root and integrate $x$. Now, we evaluate the momentum integral and take the asymptotic behavior of the integrand for $N \gg 1$. Retaining terms up to order $1/N ^{2}$ we find that the energy density per unit area, in place of Eq. (\ref{0EnergyMassless5}), is
\begin{align}
\mathcal{E} (m,L) & = - \frac{\pi ^{2}}{1440} \frac{\hbar c}{L ^{3}} \left( B _{1} + \frac{B _{2}  + B _{3} }{N ^{2}} \right) , \label{0EnergyMassiveFin}
\end{align}
where
\begin{align}
B _{1} (N, \delta) &= 120 \sum _{n = 1} ^{N-1} \left( n ^{2} + \delta ^{2} \right) ^{3/2} , \label{B1}  \\ B _{2} (N, \delta) &= - 15 \pi ^{2} \sum _{n = 1} ^{N-1} n ^{4} \left( n ^{2} + \delta ^{2} \right) ^{1/2} , \label{B2} \\ B _{3} (N, \delta) &= - 6 \pi ^{2}  \sum _{n = 1} ^{N-1}  \left( n ^{2} +\delta ^{2} \right) ^{5/2} , \label{B3}
\end{align}
and $\delta \equiv 2L / \lambda _{\mbox{\tiny C}}$. Here, $\lambda _{\mbox{\tiny C}}= h/mc$ is the Compton length of the massive field. Note that in the limit $\delta \to 0 $ the Casimir energy (\ref{0EnergyMassiveFin}) correctly reduces to that of the massless case since the $B$-coefficients (\ref{B1})-(\ref{B3}) reduce to the $A$-coefficients given by Eq. (\ref{A123}). Now we have to evaluate these partial sums. Accordingly, we will follow the same program introduced in the previous section. In the present case, we conveniently introduce the following exponential cutoff function
\begin{align}
\eta _{\mbox{\tiny E}} (n/N , \delta / N) = \exp \left\lbrace- \frac{\sqrt{n ^{2} + \delta ^{2}}}{N} \right\rbrace , \label{ExpCutoff}
\end{align}
which reduces to the one employed in the massless case for $\delta = 0$, as expected. Therefore, the partial sums appearing in the $B$-coefficients may be evaluated by exponential function regularization, with the cutoff function of Eq. (\ref{ExpCutoff}). We will first consider the unregulated series
\begin{align}
Q(s , \delta ) = \sum _{n=1} ^{N-1} \left( n ^{2} + \delta ^{2} \right) ^{s}  \label{S-PartialSum}
\end{align}
appearing in the coefficients $B _{1}$ and $B _{3}$ with $s = 3/2$ and $s = 5/2$, respectively. The regularized sum then reads 
\begin{align}
Q (s , \delta ) = \sum _{n = 1} ^{\infty} \left( n ^{2} + \delta ^{2} \right) ^{s} \eta _{\mbox{\tiny E}} (n/N , \delta / N) . 
\end{align}
By Taylor expanding the regulator and keeping terms up to order $1/N ^{2}$ we find that the partial sum (\ref{S-PartialSum}) can be approximated as
\begin{align}
Q ( s , \delta ) \simeq \;  & Z (-s , \delta ) - \frac{1}{N} Z ( - s - 1 /2 , \delta ) \notag \\ & + \frac{1}{2N ^{2}} Z ( - s - 1 , \delta ) + \mathcal{O} (N ^{-3}) ,
\end{align}
where each of these terms can be regarded as particular values of the Epstein-Hurwitz zeta function:
\begin{align}
Z ( s , \delta ) = \sum _{n = 1} ^{\infty} \left( n ^{2} + \delta ^{2} \right) ^{ - s} , \quad \mbox{Re} (s) > 1/2 \, ,  \label{Epstein-Hurwitz} 
\end{align}
which admits, however, an analytic continuation to $\mbox{Re} (s) < 1/2$. As shown in Ref. \cite{Elizalde}, the meromorphic function in the whole complex plane
\begin{align}
\!\! Z ( s , \delta ) &= - \frac{1}{2 \delta ^{2s}} + \frac{\sqrt{\pi}}{2 \delta ^{2s-1} \Gamma (s)} \notag \\ & \times \! \left\lbrace \Gamma \left( s - 1/2 \right) \! + \! 4 \sum _{n =1} ^{\infty} (\pi n \delta) ^{s - 1/2} K _{s - 1/2} (2 \pi n \delta) \right\rbrace , \label{Analytic-Zeta}
\end{align}
provides the analytic continuation of the Epstein-Hurwitz zeta function (\ref{Epstein-Hurwitz}). Here, $K _{n}$ is the $n$-th order modified Bessel function of second type \cite{Ryzhik}.

Now let us consider the unregulated series
\begin{align}
R (s , \delta ) = \sum _{n=1} ^{N-1} n ^{4} \left( n ^{2} + \delta ^{2} \right) ^{s} , \label{S-PartialSum}
\end{align}
which occurs in the coefficient $B _{2}$ with $s = 1/2$. In a similar fashion, the regularized sum
\begin{align}
R (s , \delta ) = \sum _{n=1} ^{N-1} n ^{4} \left( n ^{2} + \delta ^{2} \right) ^{s} \eta _{\mbox{\tiny E}} (n/N , \delta / N) ,
\end{align}
can be approximated (by Taylor expanding the regulator) up to order $1/N ^{2}$ as
\begin{align}
R (s , \delta ) \simeq \;  & W ( - s , \delta ) - \frac{1}{N} W ( -s - 1/2 , \delta ) \notag \\ & + \frac{1}{2N ^{2}} W ( -s-1 , \delta ) + \mathcal{O} (N ^{-3}) ,
\end{align}
where we have defined the function
\begin{align}
W ( s , \delta ) = \sum _{n = 1} ^{\infty} n ^{4} \left( n ^{2} + \delta ^{2} \right) ^{ - s}  ,  \label{ECM-function} 
\end{align}
which is absolutely convergent in the region $\mbox{Re} (s) > 0$. As usual, it can be extended to the region $\mbox{Re} (s) < 0$ by analytic continuation. Details of technical computations are relegated to Appendix \ref{Appendix}. Here we just present the final result. The function
\begin{align}
W ( s , \delta ) & = \frac{3 \sqrt{\pi}}{8 \delta ^{2s - 5} \Gamma (s)}  \Bigg\{\ \Gamma (s - 5/2)  + \frac{8}{3} \sum _{n = 1} ^{\infty}  \left( n \pi \delta \right)^{s-5/2} \notag \\ & \phantom{=} \times \left\lbrace 2 \left[ (n \pi \delta) ^2+(s-3) s+2\right] K_{s- 5/2}(2 n \pi  \delta ) \right. \notag \\ & \phantom{=} \left. + (n \pi  \delta ) (1-2 s) K_{s- 3/2}(2 n \pi  \delta )\right\rbrace  \Bigg\}\  \label{ECM-Analytic-function} 
\end{align}
is what provides the analytic continuation of (\ref{ECM-function}) to the whole complex plane. 

With the help of the results of Eqs. (\ref{Analytic-Zeta}) and (\ref{ECM-Analytic-function}), we can directly compute the functions $Q(3/2, \delta)$, $R (1/2, \delta)$ and $Q(5/2, \delta)$, appearing in the $B$-coefficients. Substituting these functions in Eq. (\ref{0EnergyMassiveFin}) we obtain an explicit expression for the energy density per unit area (up to order $1/N ^{2}$). For simplicity, we write the energy density as the sum of three terms:
\begin{align}
\mathcal{E} (m,L) = \mathcal{E} _{\mbox{\scriptsize} C} + \mathcal{K}  \, L + \mathcal{E} _{\mbox{\scriptsize P}} (m,L) . \label{TotalEnergyMassive}
\end{align}
In this expression,
\begin{align}
\mathcal{E} _{\mbox{\scriptsize} C}  = \frac{\pi ^{2}}{15} \frac{\hbar c}{\lambda _{\mbox{\tiny C}} ^{3}} \left[5 \left( 1 - \frac{2 \lambda}{\lambda _{\mbox{\tiny C}}} \right) - (10 + \pi ^{2}) \frac{\lambda ^{2}}{\lambda _{\mbox{\tiny C}} ^{2}} \right]
\end{align}
is a constant term (independent of $L$) which arises from the first term in Eq. (\ref{Analytic-Zeta}). Since we are interested in variations of the energy with respect to $L$, this constant term does not have physical significance, so it will be dropped. 

On the other hand, the first term in the curly brackets of Eqs. (\ref{Analytic-Zeta}) and (\ref{ECM-Analytic-function}) produces the second term ($\sim \mathcal{K} \, L$) in the energy density (\ref{TotalEnergyMassive}), where
\begin{align}
\mathcal{K} = - \frac{\pi ^{2}}{4} \frac{\hbar c}{\lambda _{\mbox{\tiny C}} ^{4}} \left[ 2 \, \Gamma (-2) - (10 - 3 \pi ^{2}) \, \Gamma (-3)  \frac{\lambda ^{2}}{\lambda _{\mbox{\tiny C}} ^{2}} \right] 
\end{align}
is a constant with units of pressure. As we shall see, although this term contributes to the energy density inside the cavity, it cancels out with the contributions outside the plates.

The last term in the energy density,
\begin{align}
\mathcal{E} _{\mbox{\scriptsize P}} (m,L) & = 15 \delta ^{4} \mathcal{E} _{0} (L) \sum _{n=1} ^{\infty}  \Bigg\{\ \!\!  \Big[ 12  + (\pi \delta / N) ^{2} \Big] \frac{K _{2} (2 n \pi \delta ) }{(n \pi \delta) ^{2}} \notag \\ & \phantom{=}  + \! (\pi \delta / N) ^{2} \! \left[ (n \pi \delta) ^{2} + \frac{21}{4} - \frac{15}{\pi ^{2}} \right] \! \frac{K _{3} (2 n \pi \delta ) }{(n \pi \delta) ^{3}} \Bigg\}\  , \label{EnergyMassive}
\end{align}
gives the physically relevant energy density per unit area. This becomes clear when we write the total energy density of the system, i.e.
\begin{align}
\mathcal{E} _{\mbox{\scriptsize T}} (m,L) = \mathcal{E} (m,L) + \mathcal{E} (m , L _{1} - L)  + \mathcal{E} (m , L _{2}) ,   \label{TotalEnergyFullSystem}
\end{align}
where the first term is the energy density from the field inside the plates, and the second and third terms are the contributions from the field outside the plates. By putting (\ref{TotalEnergyMassive}) into (\ref{TotalEnergyFullSystem}) it turns out that the total energy density acquires constant terms and as such they do not have physical significance. As it should be, the terms proportional to $L$ cancel out, i.e. $\mathcal{K} \, L + \mathcal{K} \,  (L _{1} - L) = \mathcal{K} \, L _{1}$, and neither they will contribute to the pressure. This leaves us with
\begin{align}
\mathcal{E} _{\mbox{\scriptsize T}} (m,L) = \mathcal{E} _{\mbox{\scriptsize P}} (m,L) + \mathcal{E} _{\mbox{\scriptsize P}} (m,L _{1} - L) \label{NewEnergy}
\end{align}
as the physically relevant contribution to the energy density. In particular, in the limit in which $L _{1} \gg L$, the last term in Eq. (\ref{NewEnergy}) is strongly suppressed with respect to the first one, and thus we get $\mathcal{E} _{\mbox{\scriptsize T}} (m,L) = \mathcal{E} _{\mbox{\scriptsize P}} (m,L) $ as the Casimir energy of the system. In the following we concentrate in the evaluation of the Casimir energy.

The summations in Eq. (\ref{EnergyMassive}) can not be calculated explicitly, therefore we will consider its asymptotic limits. For small $\delta$, which implies $mL \ll h/2c$, it is possible to use the expansion \cite{Ryzhik}
\begin{align}
(x/2) ^{\nu} K _{\nu} (x) \sim \frac{1}{2} \Gamma (\nu) + \mathcal{O} (x ^{2}), \quad \nu > 0, \quad x \ll 1
\end{align}
and the definition of the Riemann zeta function (\ref{RiemannZ}) to obtain, up to second order in $\delta$:
\begin{align}
\mathcal{E} _{\mbox{\scriptsize T}} (m,L) & \simeq \mathcal{E} _{0} (L) \; \Bigg[  1 + \frac{1}{N ^{2}} \left( \frac{\pi ^{2}}{12} - \frac{5}{21} \right)  - 15 \delta ^{2} \notag \\ & \hspace{2.7cm} + \frac{3}{4} \left( \frac{5}{3}  - \frac{\pi ^{2}}{4} \right) \frac{\delta ^{2}}{N ^{2}}  \Bigg] . \label{EpolyMassiveSmall}
\end{align}
In this expression we identify the first two terms with the Casimir energy for a massless polymer scalar field, which is the result given by Eq. (\ref{0EnergyMasslessFIN}). The third term ($\propto \delta ^{2}$) corresponds to the usual small correction term due to the mass of the field. The last term, where the mass of the field and the minimal length of the theory occur, is even a smaller correction term. The Casimir pressure between the plates is given by $\mathcal{P} _{\mbox{\scriptsize T}}  = - \frac{1}{2 \lambda} \left[ \mathcal{E} _{\mbox{\scriptsize T}} (m, L+ \lambda) - \mathcal{E} _{\mbox{\scriptsize T}} (m, L - \lambda) \right]$. Keeping terms up to order $1/N ^{2}$ we obtain
\begin{align}
\mathcal{P} _{\mbox{\scriptsize T}} (m,L) &= - \frac{\pi ^{2}}{480} \frac{\hbar c}{L ^{4}} \left[ 1 + \frac{5 \lambda ^{2}}{9 L ^{2}} \left( \frac{\pi ^{2}}{4} - \frac{37}{7} \right) + \frac{20 L ^{2}}{\lambda _{\mbox{\tiny C}} ^{2}} \right. \notag \\ & \hspace{2cm} + \left. \frac{15 \lambda ^{2}}{\lambda _{\mbox{\tiny C}} ^{2}} \left( 1 + \frac{\pi ^{2}}{20} \right) \right] , \label{ForceMassive}
\end{align}
which reduces to the massless result given by Eq. (\ref{ForceMasslessFIN}) when $\lambda _{\mbox{\tiny C}} \to \infty$. 

On the other hand, if we take $\delta \gg 1$ in Eq. (\ref{EnergyMassive}), which implies $mL \gg h/2c$, the asymptotic expansion of the modified Bessel function \cite{Ryzhik}
\begin{align}
K _{\nu} (x) \sim \sqrt{\frac{\pi}{2x}} e ^{-x} , \quad \nu > 0, \quad x \gg 1
\end{align}
together with the definition of the polylogarithm function 
\begin{align}
\mbox{Li} _{s} (z) = \sum _{k = 1} ^{\infty} \frac{z ^{k}}{k ^{s}}
\end{align}
gives
\begin{align}
\mathcal{E} _{\mbox{\scriptsize T}} (m,L) & = 90 \delta ^{3/2}  \mathcal{E} _{0} (L)  \Bigg\{\ \!\! \mbox{Li} _{5/2} (\mu) + \frac{\pi ^{2} \delta ^{2}}{12 N ^{2}} \Bigg[ \mbox{Li} _{5/2} (\mu)   \notag  \\ & \phantom{=} + \pi \delta \, \mbox{Li} _{3/2} (\mu) + \frac{1}{\pi \delta} \left( \frac{21}{4} - \frac{15}{\pi ^{4}} \right) \mbox{Li} _{7/2} (\mu) \Bigg] \! \Bigg\}\ \! ,
\end{align}
where $\mu = e ^{-2 \pi \delta}$. The leading contribution for $\delta \gg 1$ is then
\begin{align}
\mathcal{E} _{\mbox{\scriptsize T}} (m,L) & = - \frac{\hbar c}{16} \left( \frac{2}{\lambda _{\mbox{\tiny C}} L} \right) ^{3/2} e ^{-4 \pi L  / \lambda _{\mbox{\tiny C}}}  \Bigg\{\ 1 + \frac{\pi ^{2} \lambda ^{2}}{3 \lambda _{\mbox{\tiny C}} ^{2}}  \notag  \\ & \phantom{=} \times  \left[ 1 + \frac{2 \pi L}{\lambda _{\mbox{\tiny C}}}+ \left( \frac{21}{4} - \frac{15}{\pi ^{4}} \right) \frac{\lambda _{\mbox{\tiny C}}}{2 \pi L} \right] \Bigg\}\  .
\end{align}
This expression implies that the energy coming from the lower modes is dominated by $m$ and does not depend on $L$ so strongly, as in the previous case ($\delta \ll 1$).

\section{Summary and conclusions} \label{ConclusionSection}

In this paper we have analyzed the Casimir effect associated with a polymer-quantized scalar field confined between two parallel conducting plates separated by a distance $L$ along a given direction. An important feature about the boundary conditions is that in quantum models with a minimal length, such as the polymer theory, there is a finite number of modes allowed between the plates, $n _{\mbox{\scriptsize max}} = N = L/ \lambda$. Also, the wavelength cannot take arbitrary values but has a minimum value $\lambda$. This in turn implies that the momenta in the plane parallel to the plates have a maximum value given by $\pi \hbar / \lambda$. So there are natural cutoff values  and the Casimir energy does not need to be regularized, as opposed to standard quantum field theory calculations. The zero-point energy was calculated by summing over the modes. The expression for the Casimir energy density inside the plates was found to be
\begin{align}
\mathcal{E} (m,L) = \sum _{n = 1} ^{n _{\mbox{\tiny max}}} \int _{\mathcal{D} \subseteq \mathbb{R} ^{2}}   E _{\lambda , n} (\vec{k}) \, \frac{d ^{2} \vec{k}}{(2 \pi) ^{2}} ,
\end{align}
where $\mathcal{D}$ is the rectangular domain defined by Eq. (\ref{RegionD}), and $E _{\lambda , n} (\vec{k})$ is the energy spectrum for a massive polymer scalar field confined between the plates. The finite size of the region $\mathcal{D}$ as well as the truncated sum are direct consequences of the minimal length of the theory. Of course, in the limit $\lambda / L \to 0$, $n _{\mbox{\tiny max}} \to \infty$ and $\mathcal{D} \to \mathbb{R} ^{2}$. To evaluate this expression, we have safely assumed $\lambda \ll L$, since there is no evidence of the spatial discreteness. This allowed us to express the Casimir energy as an expansion in powers of $1/N$. We have computed the leading order correction in $1/N$ to the Casimir energy in order to show analytical results. In order to consider the contributions from the field inside and outside the plates, we introduce two auxiliary plates at $z = - L _{2}$ and $z = L _{1} \gg L$, such that the total energy of the system has the form $\mathcal{E} (m,L) + \mathcal{E} (m,L _{1} - L) + \mathcal{E} (m,L _{2})$. At the end of the calculations we take the limit $L _{1} \gg L$, which is the one that leaves us with the stress (per unit area) between the plates at $z = 0$ and $z = L$. The Casimir pressure for a massless polymer scalar field is shown in Fig. \ref{CasForceMassless}. Using the experimental results reported in Ref. \cite{Bressi}, we have obtained an upper bound for the minimum length, i.e. $\lambda \sim 10 ^{-7}$m, which is far from the one expected if we think $\lambda$ as arising from a more fundamental quantum theory of gravity.

We have also computed the Casimir energy and pressure for a massive polymer scalar field. In order to obtain analytical results, we first considered the small mass limit ($mL \ll h / 2c$). In this case the correction term due to the polymer length scale is of the same sign than the continuous case. In other words, the Casimir pressure $\mathcal{P}$ is larger than $\mathcal{P} _{0} = - \frac{\pi ^{2}}{480}  \frac{\hbar c}{L ^{4}}$, which is the usual pressure for a massless scalar field satisfying the Klein-Gordon equation \cite{Milton, Bordag}. In the large mass limit ($mL \gg h / 2c$), we found that the Casimir energy is dominated by the mass and does not depend on $L$ so strongly as in the previous case.

Now we close by discussing two scenarios where the Casimir effect has been studied for i) a Lorentz-violating scalar field and ii) theories with minimal length (within GUPs), and which can be compared with the results of the present study.

On the one hand, in Refs. \cite{Petrov1, Petrov2}, the Casimir effect was studied for a Lorentz-violating scalar field theory. There, the authors introduced a theoretical model for the Klein-Gordon field which incorporates Lorentz symmetry violation through a fixed four-vector $\lambda u ^{\mu}$. After imposing the appropriate boundary conditions to the field at the plates, they computed the Casimir energy by summing over the modes and obtain analytical expressions (under the assumption $\lambda \ll 1$). When $u ^{\mu}$ is timelike, they found that the influence of the Lorentz-symmetry breaking parameter upon the Casimir energy consists only of a multiplicative factor. The case for a spacelike $u ^{\mu}$ is more complex. When $u ^{\mu}$ points in the direction perpendicular to the plates, the Casimir energy acquires a complicated dependence on $\lambda$. It is worth mentioning that the summation over the modes allowed between the plates and the integrals over the transverse momenta are performed as usual, up to infinity and in the whole $\mathbb{R} ^{2}$, respectively. So they used the Abel-Plana formula to regularize the vacuum energy. Note that in the model of Refs. \cite{Petrov1, Petrov2}, the Lorentz-breaking term is inspired in the so-called Standard Model Extension \cite{Kostelecky}, which is an effective field theory which incorporates Lorentz-breaking but gauge-invariant terms, while in the present study, Lorentz symmetry is naturally broken due to the spatial discreteness of the polymer theory. Therefore, these studies have deep different physical origins.

On the other hand, the Casimir effect has been studied in minimal length theories based on a Generalized Uncertainty Principle \cite{Ulrich, Nouicer, Frasino}. There, the minimal length is implemented by adding corrections to the usual position-momentum commutation relation in this way: $[x _{i} , p _{j}] = i \hbar [f (p ^{2}) \delta _{ij} + g (p ^{2}) p _{i} p _{j}]$, where the functions $f (p ^{2})$ and $g (p ^{2})$ are not completely arbitrary. Different choices of these functions leads to different quantum theories or GUPs. As an example, take $f (p ^{2})= 1 + \beta p ^{2} $ and $g (p ^{2}) = 0$. This yields to the modified uncertainty relation $\Delta x \Delta p \geq \frac{\hbar}{2} [1 + \beta (\Delta p) ^{2} + \gamma]$ with $\beta , \gamma > 0$. Clearly, there is a finite minimal uncertainty $\Delta x _{0} = \hbar \sqrt{\beta}$, which is regarded as the fundamental length of the theory. This case has some resemblance with the one presented in this work in the sense that the number of discrete modes allowed between the plates and the momentum parallel to the plates, acquire natural cutoff values due to the minimal uncertainty $\Delta x _{0}$. Despite this, theories arising from the GUPs are still continuous models. In our case, the spatial discreteness produces difference equations (in coordinate representation) instead of differential equations, and the minimum length scale is uniquely defined.

\acknowledgments

A.M.R. acknowledges support from DGAPA-UNAM project IA101320. C.A.E. is supported by UNAM-DGAPA postdoctoral fellowship and the project PAPIIT IN111518. We also are indebted to the reviewer for his/her valuable comments and suggestions to improve the quality of the paper.

\appendix

\section{Analytic continuation of $W ( s , \delta )$} \label{Appendix}

The analytic continuation of the function $W ( s , \delta )$ can be done in the following way. Using the integral representation of the gamma function (\ref{Exp-Int}) we get
\begin{align}
W ( s , \delta ) = \frac{1}{\Gamma (s)}  \int _{0} ^{\infty} t ^{s-1} e ^{-t  \delta ^{2}} \; \frac{d ^{2} S (t)}{d t ^{2}} \; dt ,
\end{align}
where $S (t)$ is the analytic function
\begin{align}
S (t) = \sum _{n = 1} ^{\infty} e ^{-t n ^{2} } . \label{S-func-App}
\end{align}
Integrating by parts twice one finds
\begin{align}
W ( s , \delta ) = \frac{1}{\Gamma (s)}  \int _{0} ^{\infty} \Big[ \delta ^{4}  t ^{s-1} - 2 \delta ^{2} (s-1)  t ^{s-2} \notag \\ + (s-1) (s-2)  t ^{s-3} \Big] e ^{-t  \delta ^{2}} S (t) \; dt . \label{S-int}
\end{align}
To move forward, we use the identity
\begin{align}
S (t) = - \frac{1}{2} + \frac{1}{2} \sqrt{\frac{\pi}{t}} + \sqrt{\frac{\pi}{t}} \; S \left( \pi ^{2} / t \right) , \label{S-identity}
\end{align}
which can be directly derived from the definition (\ref{S-func-App}). Inserting the identity (\ref{S-identity}) into the integral in (\ref{S-int}) and after performing simple exponential integrals we get
\begin{align}
W ( s , \delta ) &= \frac{3 \sqrt{\pi}}{8 \Gamma (s)} \delta ^{5-2s} \Gamma (s - 5/2) + \frac{\sqrt{\pi}}{ \Gamma (s)}  \int _{0} ^{\infty} t ^{-1/2} e ^{-t  \delta ^{2}} \notag \\ & \times \! \big[ \delta ^{4}  t ^{s-1} \! - 2 \delta ^{2} (s-1)  t ^{s-2} \! + \! (s-1) (s-2)  t ^{s-3} \big] \notag \\ & \times  S \left( \pi ^{2} / t \right) dt .
\end{align}
Substituting the function $S \left( \pi ^{2} / t \right)$ from its definition (\ref{S-func-App}) and using the following integral representation of the modified Bessel function
\begin{align}
\int _{0} ^{\infty} x ^{\nu - 1} e ^{- \frac{a}{x} - b x} dx = 2 \left( \frac{a}{b} \right) ^{\nu /2} K _{\nu} \left( 2 \sqrt{ab} \right) ,
\end{align}
we finally obtain
\begin{align}
W ( s , \delta ) & = \frac{3 \sqrt{\pi}}{8 \delta ^{2s - 5} \Gamma (s)}  \Bigg\{\ \Gamma (s - 5/2)  + \frac{8}{3} \sum _{n = 1} ^{\infty}  \left( n \pi \delta \right)^{s-5/2} \notag \\ & \phantom{=} \times \left\lbrace 2 \left[ (n \pi \delta) ^2+(s-3) s+2\right] K_{s-5/2}(2 n \pi  \delta ) \right. \notag \\ & \phantom{=} \left. + n \pi  \delta (1-2 s) K_{s-3/2}(2 n \pi  \delta )\right\rbrace  \Bigg\}\ ,
\end{align}
which does in fact provide the analytic continuation of the function (\ref{ECM-function}).

\end{document}